\documentclass[aps,prb,twocolumn,showpacs]{revtex4-1}
\usepackage{graphicx}
\usepackage{epsfig}
\usepackage{times}

\begin{document}

\title{Thermally activated in-plane magnetization rotation induced by  spin torque}
\author{L. Chotorlishvili$^1$, Z. Toklikishvili$^2$, A. Sukhov$^1$, P. P. Horley$^3$, \\ V. K.~Dugaev$^{4,5}$, V. R. Vieira$^5$, S. Trimper$^1$
and J. Berakdar$^1$} \affiliation{
$^1$Institut f\"ur Physik, Martin-Luther-Universit\"at Halle-Wittenberg, Heinrich-Damerow-Str. 4, 06120 Halle, Germany\\
$^2$Physics Department of the Tbilisi State University, Chavchavadze av. 3, 0128, Tbilisi, Georgia\\
$^3$Centro de Investigacion en Materiales Avanzados (CIMAV S.C.), Chihuahua/Monterrey\\
$^4$Department of Physics, Rzesz\'{o}w University of Technology Al. Powstanc\'{o}w Warszawy 6, 35-959 Rzesz\'{o}w, Poland\\
$^5$Department of Physics and CFIF, Instituto Superior T\'ecnico, TU Lisbon, Av. Rovisco Pais, 1049-001 Lisbon, Portugal}

\begin{abstract}
We study the role of thermal fluctuations on the spin dynamics of a thin permalloy film
with a  focus on the behavior of spin torque and find that the thermally assisted
 spin torque results in new aspects of   the magnetization dynamics. In particular, we uncover
the formation of a finite,  spin torque-induced,   in-plane magnetization component.
The orientation of the in-plane magnetization vector depends on the temperature and the spin-torque coupling.
We investigate and illustrate that  the variation of the temperature leads to a thermally-induced rotation
of the in-plane magnetization.
\end{abstract}
\date{\today }

\maketitle

\section{Introduction}

We are witnessing  a growing body of research on various phenomena related to the transfer of
angular momentum by means of an electric current \cite{torque_rev}. The fact that the electron
current carries/transfers spin is well-known,  the interest to
this phenomenon was fueled however by  the experimental demonstrations that the
electric current can strongly affect the magnetization dynamics in nanostructures
\cite{Yamaguchi,Chiba,Saitoh,Klaui} with important consequences for technological applications
such as steering
magnetic domain walls \cite{10} and vortices, conception of high-frequency electrical oscillators \cite{11,12}, and
the  magnetization reversal in magnetic layers via exerting spin torque \cite{9}.
The latter is achieved  by a spin-polarized
charge current and has been demonstrated for various magnetic nanostructures
\cite{Taniguchi,d'Aquino,Eltschka,Schieback,Balaz} and magnetic tunnel junctions
\cite{Sun}.
While several  microscopic mechanisms
relevant for various nanosystems have been discussed,  on the macroscopic level
the effect of the spin-polarized current can also be  described
by the well-established macroscopic  Landau-Lifshitz-Gilbert (LLG) equation \cite{jap_thermal} upon
including  the appropriate spin-torque terms.

Another important factor, which can  influence substantially the
magnetization switching in nanostructures, is the effect of thermal
fluctuations \cite{Brown} (here we refer to the very comprehensive recent overview \cite{jap_thermal} and references therein for the
details of  the
well-studied finite-temperature spin dynamics).
 This effect can be captured by including fluctuating Langevin fields into the LLG equation.
Following the standard protocol \cite{Brown}, the magnetization trajectory
can be identified as the average over the ensemble of noninteracting
nanoparticles and described by the Fokker-Planck (FP)
equation.
%We show below, that the thermal in nature of the magnetization dynamics
%in the presence of  the spin torque
%term and may lead, as shown below, to rich and highly
%nontrivial dynamical effects in the magnetization dynamics described
%by the LLG equation.

In this paper we consider the combined influence of thermal fluctuations and the
spin torque terms previously derived
for the case of  the current-induced motion of a
magnetic domain wall in a quasi-one-dimensional ferromagnet with
easy-axis and easy-plane anisotropies \cite{Dugaev}. We
show that such a torque term leads to an interesting physical
phenomenon of thermally activated in-plane magnetization rotation.

We will show that in case of spin torque exerted by spin polarized current, orientation of the in-plane magnetization can be easily switched and controlled by thermal heating or thermal cooling of the system. Discovered effect may have promising applications based on controlling magnetization dynamics in nanostructures. A key issue in our result is the ratio between the thermal activation energy and the Zeeman energy of the magnetization vector in the external driving magnetic field. This means from the experimental viewpoint that our theoretical proposal can be easily implemented by tuning the amplitude of the external driving magnetic field. Thus, the in-plane magnetization vector can be controlled and switched by out-of-plane external magnetic field.

 To obtain analytical solutions of the FP equation we
develop a perturbation approach which substantially differs from the previously discussed
methods \cite{Evans,Usadel,Fuchs,Kalmykov,Titov,Coffey,Fannin}. The advantage
of our approach is that it allows for obtaining some analytical solutions with
high accuracy in arbitrary order of the perturbation theory.

\section{Model}

The finite temperature magnetization dynamics in a thin ferromagnetic layer in the
presence of a spin torque and an external magnetic field can be
 described by the following stochastic LLG equation
\cite{Dugaev,Brown,Sementsov}:
\begin{eqnarray}\label{eq.1}
&&\frac{d{\bf M}}{dt}=\gamma_{e}{\bf M}\times \big({\bf H}_{eff}+{\bf h}(t)\big) \nonumber
\\&& -\gamma_{e}\lambda {\bf M}\times \big[{\bf M}\times \big({\bf H}_{eff}+{\bf h}(t)\big)\big] \nonumber
\\&& + b{\bf M}\times {\bf s}
+ a{\bf M}\times \big({\bf M}\times {\bf s}\big).
\end{eqnarray}
 Here $H_{eff}$ is the effective magnetic field and $h(t)$ is the random
 Langevin magnetic field related to the thermal fluctuations, $a,b$ are the
 Slonczewski spin torque constants, $\gamma_{e}$ is the gyromagnetic ratio for electrons, and  $\lambda$ is
 a phenomenological (Gilbert) damping  constant. For
convenience we  introduce  dimensionless quantities.
Thus we deal with a normalized magnetization vector
$|{\bf M}|=1$, a dimensionless (rescaled) damping
$\lambda\rightarrow \lambda /|{\bf M}|$, the torque constants
$a/\omega_{0}\rightarrow\varepsilon a$, and
$b/\omega_{0}\rightarrow\varepsilon b$.  The dimensionless time
$t\rightarrow \omega_{0}t$ is defined through the frequency of the Larmor precession
 in the effective field $\omega_{0}=\gamma|{\bf
H}_{eff}|$. The anisotropy field for the ferromagnetic system can
be evaluated for  thin film alloys of the  permalloy class (Fe-Ni,
Fe-Co-Ni) by using the formula \cite{Rado,Ugulava}
$\beta_{A}=2K_{1}/M_{s}$, where $K_{1}$ is the anisotropy
coefficient and $M_{s}$ is the saturation magnetization. In
particular, for a thin film \cite{Sementsov,Ugulava, Kikoin}
Fe$_{25}$Co$_{25}$-Ni$_{50}$ the saturation  magnetization of the
film is of the order of $M_{s}\approx1025 $~G, the anisotropy
constant $K_{1}\approx 4\times 10^{3}$~erg/cm$^{3}$, the
anisotropy field $\beta_{A}\approx7.8$~Oe, and the anisotropy
field in units of the frequency is
$\omega_{p}=\gamma_{e}\beta_{A}\approx0.138\times10^{9}$~Hz
$(\gamma_{e}=1.755\times 10^{7}$~Oe$^{-1}$c$^{-1}$, while the
Zeeman frequency in the reasonable strong external magnetic field
is $\omega_{0}=\gamma_{e}|{\bf
H}_{0}|\approx17\times(10^2\div10^4)$~MHz $(|{\bf H}_{0}|\approx
10^{2}\div10^{4}$~Oe). Since $\omega_{0}>>\omega_{p}$ we conclude
that for the Fe-Co-Ni alloy, the dominating factor is the external
magnetic field ${\bf H}_{eff}=(0,~0,~H_{0})$.

The components of the Langevin field
$h_{\alpha},~\alpha=x,y,z$ obey the following correlation relations:
\begin{eqnarray} \label{eq.2}
   &\langle {\bf h}(t)\rangle=0,&  \nonumber\\
   &\langle h_{\alpha}(t)h_{\beta}(t')\rangle=2\lambda T\delta_{\alpha\beta}\, \delta(t-t'),&
\end{eqnarray}
where the averaging ($\langle \cdots \rangle$) is performed  over all possible realizations of the random field  ${\bf h}(t)$.
For the derivation of the stochastic Fokker-Planck equation we follow Ref.~\onlinecite{Garanin} and use
the functional integration method in order to average the dynamics over all possible
realizations of random noise field.
As shown in Ref.~\onlinecite{Garanin}, this method is quite general and straightforward,
and for the case of a small coupling between the system and the bath it recovers previously
obtained results\cite{Brown}.

We define the distribution function in the following form
\begin{equation}\label{eq.3}
    f({\bf N},t)=\langle\pi(t,[h])\rangle_{h},~~~\pi(t,[h])
    =\delta({\bf N}-{\bf M}(t)).
\end{equation}
Here ${\bf N}$ is the unit vector on the sphere, and we assume that the random field ${\bf h}(t)$
stands for a Gaussian noise with the associated  functional:
\begin{equation}\label{eq.4}
F\big[{\bf h}(t)\big]=\frac{1}{Z_{h}}
\exp\bigg[-\frac{1}{2g}\int\limits_{-\infty}^{+\infty}d\tau \, {\bf h}^{2}(\tau)\bigg].
\end{equation}
Here $Z_{h}=\int D{\bf h}\, F$ is the normalization factor and  $\int D{\bf h}$
denotes the functional integration over all possible realizations of the random field
${\bf h}(t)$, and $g=2\lambda T$.
Note that for convenience we measure the temperature in units of the Larmor frequency
$\omega_{0}=\gamma|{\bf H}_{eff}|$. Therefore, the dimensionless temperature is
defined via the expression $T\rightarrow k_{B}T/\omega_{0}\hbar $.

Taking into account the relations \cite{Garanin}:
\begin{eqnarray}\label{eq.5}
   &&\frac{\delta h_{\alpha}(\tau)}{\delta h_{\beta}(t)}
   =\delta_{\alpha\beta}\, \delta(\tau-t),  \nonumber
\\
   &&\int D{\bf h}\frac{\delta^{n}F[{\bf h}]}
   {\delta h_{\alpha_{1}}(t_{1})\delta h_{\alpha_{2}}(t_{2})\ldots \delta h_{\alpha_{n}}(t_{n})}=0,
\\
   &&\frac{d \pi}{dt}=-\frac{\partial \pi}{\partial {\bf N}}\frac{d {\bf M}}{dt},  \nonumber
\end{eqnarray}
and following the standard procedure \cite{Garanin}, we deduce from Eq.(\ref{eq.1}) the
following FP equation:
\begin{eqnarray} \label{eq.6}
   \frac{\partial f}{\partial t}
   =-\frac{\partial}{\partial{\bf N}}
  \bigg\{-\bf N\times {\bf H}_{eff}
  +\lambda {\bf N}\times {\bf N}\times {\bf H}_{eff}+
  \nonumber\\
  +\varepsilon b\, {\bf N}\times {\bf s}
  +\varepsilon a \, {\bf N}\times {\bf N}\times {\bf s}
  -\lambda   T\, {\bf N}\times {\bf N}\times \frac{\partial}{\partial {\bf
  N}}\bigg\}f .
\end{eqnarray}
Solving for such a time-dependent FP equation is a difficult problem
even in the absence of spin torque terms
\cite{Kalmykov,Titov,Coffey,Fannin}. In the presence of spin
torque, the analytical consideration of the non-stationary FP equation
becomes intractable. To proceed further we consider a
particular configuration of the spin torque ${\bf s}=(s,0,0)$
\cite{Dugaev,Balaz} and the driving external field
${\bf H}_{eff}=(0,0,H_{0})$ terms. Here for convenience, the amplitude of the
renormalized magnetic field is set to one.  We will look for the
perturbation solution of  Eq.(\ref{eq.6}) and consider the case
$\varepsilon =1/\omega_{0}\ll 1$ ($\omega_{0}$ is the Larmor
precession frequency in the external constant magnetic field) as
a small parameter of the theory and look for a stationary
solution of Eq.(\ref{eq.6}) in form
\begin{equation}\label{eq.7}
    f=C\exp\bigg[\frac{1}{T}\big({\bf N}\cdot{\bf H_{eff}})+\varepsilon \psi({\bf N})\ldots\bigg] .
\end{equation}
Here $\psi(\bf N)$ is a function of the vector $\bf N$. A
zeroth-order solution $f_{0}=C\exp\bigg[\frac{1}{T}\big({\bf
N}\cdot{\bf H_{eff}}\big)\bigg]$ corresponds to the solution in
the absence of the spin torque. In the stationary case, inserting
(\ref{eq.7}) in (\ref{eq.6}) and after straightforward
calculations we obtain
\begin{eqnarray}\label{eq.8}
    &&f=C\exp\bigg[\frac{1}{T}({\bf N}\cdot{\bf H_{eff}})+\varepsilon\frac{a}{\lambda T}({\bf N}\cdot{\bf s}) \nonumber \\
    &&+\frac{\varepsilon b}{2\lambda T^{2}}{\bf N}\cdot ({\bf s}\times
{\bf H_{eff}})+\ldots\bigg)+\ldots\bigg].
\end{eqnarray}
Here in (\ref{eq.8}) we assume  a high temperature limit,
$\beta=1/T=k_{B}/\gamma_{e}H_{0}\hbar <<1$,
and therefore we can neglect higher order terms in the inverse temperature.
For convenience, in the intermediate equations in what follows we set
$\omega_{0}=\gamma_{e}H_{0}=1$, $k_{B}=1$, $\hbar=1$.
As we show below, the values of the temperature defines the limits of the application
of the perturbation theory.
We also neglect the higher order terms that are proportional to the small parameter $\varepsilon$.

\section{Observable quantities}

Using the distribution function (\ref{eq.8}) we can evaluate the mean values of the
components of the magnetization vector using the following parametrization:
$M_{x}=\sin\theta\cos\varphi$, $M_{y}=\sin\theta\sin\varphi$, $M_{z}=\cos\theta$, $0\leq\theta\leq\pi$, $0\leq\varphi\leq2\pi$.
In the absence of the spin torque, the distribution function takes on the following form
\begin{equation}\label{eq.9}
    dw(\theta,\varphi)=f(\theta)\, d\Omega=Z^{-1}(\beta H_{0})\, \exp(\beta H_{0} \cos\theta)\, d\Omega .
\end{equation}
Here
\begin{equation}\label{eq.10}
    Z(\beta H_{0})=\int \exp (\beta H_{0}\cos\theta)\, d\Omega=\frac{4\pi}{\beta H_{0}}\sinh(\beta H_{0}) ,
\end{equation}
is the partition function and $dw(\theta,\varphi)$ defines the probability that the
magnetization vector $\vec{M}$  is oriented within a solid angle of the width
$d\Omega=\sin\theta \, d\theta \, d\varphi$.
Taking into account (\ref{eq.9}), (\ref{eq.10}) we find  $\overline{M_{x}}=\overline{M_{y}}=0$ and
\begin{equation}\label{eq.11}
    \overline{M_{z}}
    %=\bigg(\coth(\beta H)-\frac{1}{\beta H}\bigg)
    =L\big(H_{0}/T\big) ,
\end{equation}
where $L(x)=\coth(x)-\frac{1}{x}$ is the Langevin function.

In the case of the high temperature  limit $H_{0}/T<1$, that means
for $T>\gamma_{e}H_{0}\hbar /k_{B}$, we have
$\overline{M_{z}}\approx \gamma_{e}H_{0}\hbar /3k_{B}T$. In the
case of low temperatures, $T<\gamma_{e}H_{0}\hbar /k_{B}$, we have
$\overline{M_{z}}=M=1$. For the square components of the
magnetization we have
\begin{eqnarray}\label{eq.12}
   \overline{M_{x}^{2}}=\overline{M_{y}^{2}}
   =\frac{L(\beta H_{0})}{\beta H_{0}},\;
   \overline{M_{z}^{2}}=1-\frac{2L(\beta H_{0})}{\beta H_{0}}.
\end{eqnarray}
We see that Eq.(\ref{eq.12}) conserves the magnetization vector
$\overline{M_{x}^{2}}+\overline{M_{y}^{2}}+\overline{M_{z}^{2}}=1$.
For the dispersion we have
\begin{eqnarray}\label{eq.13}
   &\overline{(\Delta M_{i})^{2}}=\overline{(M_{i}-\bar{M}_{i})^{2}},\, i=x,y,z,&  \nonumber \\
   &\overline{(\Delta M_{x})^{2}}=\overline{(\Delta M_{y})^{2}}=\frac{T}{H_{0}} L(H_{0}/T),&  \\
   &\overline{(\Delta M_{x})^{2}}=1-\frac{2T}{H_{0}}L(H_{0}/T)-L^{2}\big(H_{0}/T\big). & \nonumber
\end{eqnarray}
By using the explicit form of solutions (\ref{eq.11}) and partition function (\ref{eq.10})
we can evaluate the mean energy of the system:
\begin{equation}\label{eq.14}
    \bar{U}=-\frac{\partial}{\partial \beta}\ln\big(Z(\beta H)\big)=-\gamma_{e}H_{0}\hbar L\bigg(\frac{\gamma_{e}H_{0}\hbar}{k_{B}T}\bigg),
\end{equation}
and the heat capacity
\begin{equation}\label{eq.15}
    C_{V}=\bigg(\frac{\partial \bar{U}}{\partial T}\bigg)_{V}=k_{B}\left[1-\frac{(4\pi)^{2}}{Z^{2}\bigg(\frac{\gamma_{e}H_{0}\hbar}{k_{B}T}\bigg)}\right].
\end{equation}
If the spin torque terms are taken into account  the results are changed.
The distribution function takes the form
\begin{eqnarray}
   &&f=C\exp\big[\alpha\cos \theta+\varepsilon \delta \sin \theta \cos\varphi-\varepsilon \eta\sin\theta \sin \varphi\big], \nonumber \\
  &&\alpha=\beta H_{0},~~~\delta=\frac{as}{\lambda T},~~~\eta=\frac{bs}{2 \lambda T^{2}}.
\end{eqnarray}
The expressions for magnetization components in this case are quite involved and are
presented in the appendix.For the particular case of
$\varepsilon^{2}(\delta^{2}+\eta^{2})/2<<1$, i.e.
for
\begin{equation}\label{eq.17}
\bigg(\frac{1}{T}\bigg)^{2}<2\sqrt{\bigg(\frac{a}{b}\bigg)^{4}
+\frac{2\omega_{0}^{2}\lambda^{2}}{b^{2}s^{2}}}-2\bigg(\frac{a}{b}\bigg)^{2},
\end{equation}
the expressions for the mean components of the magnetization vector simplifies to
\begin{eqnarray}\label{eq.18}
   &&\overline{M_{x}}(H_{0},T)\approx -\frac{as}{\lambda H_{0} \omega_{0}}L(H_{0}/T),\nonumber  \\
   &&\overline{M_{y}}(H_{0},T)\approx +\frac{bs}{2\lambda\omega_{0}}\frac{1}{T H_{0}}L(H_{0}/T),  \\
   &&\overline{M_{z}}(H_{0},T)\approx L(H_{0}/T)-\frac{1}{2\omega_{0}^{2}}
   \bigg(\frac{a^2s^2}{\lambda^2 H_{0}^{2}}+\frac{b^2s^3}{4\lambda^2}\frac{1}{H_{0}^{2}T^{2}}\bigg) \nonumber \\
   && \times \bigg(3L(H_{0}/T)+\frac{H_{0}}{T}L^{2}(H_{0}/T)-\frac{H_{0}}{T}\bigg),\nonumber
\end{eqnarray}
and
\begin{eqnarray}\label{eq.19}
   &&\overline{M_{x}^{2}}(\omega_{0},H_{0},T)\approx \frac{T}{H_{0}}L(H_{0}/T)\nonumber \\
   &&-\frac{1}{2\omega_{0}^{2}}\bigg(\frac{a^2s^2}
   {\lambda^2 H_{0}^{2}}+\frac{b^2s^2}{4\lambda^2}\frac{1}{H_{0}^{2}T^2}
   \bigg)L^{2}(H_{0}/T)  \nonumber\\
   &&+\frac{1}{2\omega_{0}^{2}}\bigg(\frac{3a^2s^2}
   {\lambda^2H_{0}^{2}}+\frac{b^2s^2}{4\lambda^2}\frac{1}{T^{2}H_{0}^{2}}\bigg)
   \bigg(1-\frac{3T}{H_{0}}L(H_{0}/T)\bigg),\nonumber\\
  &&\overline{M_{y}^{2}}(\omega_{0},H_{0},T)\approx \frac{T}{H_{0}}L(H_{0}/T)\nonumber \\
  &&-\frac{1}{2\omega_{0}^{2}}\bigg(\frac{a^2s^2}{\lambda^2 H_{0}^{2}}
  +\frac{b^2s^2}{4\lambda^2}\frac{1}{H_{0}^{2}T^2}\bigg)L^{2}(H_{0}/T)\\
  &&+\frac{1}{2\omega_{0}^{2}}\bigg(\frac{a^2s^2}{\lambda^2 H_{0}^{2}}
  +\frac{3b^2s^2}{4\lambda^2}\frac{1}{H_{0}^2 T^{2}}\bigg)\bigg(1-\frac{3T}{H_{0}}L(H_{0}/T)\bigg),\nonumber\\
  &&\overline{M_{z}^{2}}(\omega_{0},H_{0},T)\approx\bigg(1- \frac{2T}{H_{0}}L(H_{0}/T)\bigg) \nonumber \\
  &&+\frac{1}{\omega_{0}^{2}}\bigg(\frac{a^2s^2}{\lambda^2 H_{0}^{2}}+\frac{b^2s^2}
  {4\lambda^2}\frac{1}{H_{0}^{2} T^2}\bigg)L^{2}(H_{0}/T)\nonumber \\
  && -\frac{2}{\omega_{0}^{2}}\bigg(\frac{a^2s^2}{\lambda^2 H_{0}^{2}}+\frac{b^2s^2}
  {4\lambda^2}\frac{1}{H_{0}^2 T^{2}}\bigg)\bigg(1-\frac{3T}{H_{0}}L(H_{0}/T)\bigg).\nonumber
\end{eqnarray}
From Eqs.(\ref{eq.19}) it is easy to see that the normalization condition holds,
$M^{2}=1.$ Equation (\ref{eq.17}) defines the minimum values of the
temperature, for which the solutions (\ref{eq.18}), (\ref{eq.19}) are still
valid. In particular, taking into account that
$\omega_{0}/bs\gg 1$, from (\ref{eq.17}) we obtain
\begin{equation}\label{eq.20}
T>T_{cr},~~~~T_{cr}\approx\frac{\hbar}{k_{B}}\sqrt{\frac{\omega_{0}bs}{3\lambda}}.
\end{equation}
Equation (\ref{eq.20}) shows that the temperature, above which our approach is valid,
increases with the amplitudes of external field
$\omega_{0}=\gamma_{e}H$ or of the torque $bs$.
The meaning of (\ref{eq.18}) is straightforward.  The torque leads to a formation of
transversal components
$\overline{M_{x,y}}(\omega_{0},H_{0},T)$ while the external field tries to align the
magnetization along the $z$ axis.

Taking typical values of the parameters for the thin film Fe$_{25}$-Co$_{25}$-Ni$_{50}$ such as:
 $\omega_{0}=\gamma _{e}|H_{0}| \approx 17 \times 10^{4}$MHz, $|H_{0}|=10^{4}$Oe, $\lambda=10^{-4}$
for the maximal value of the critical threshold temperature $T_{cr}$ we have $T_{cr}<70K$.

In the limit of a strong field and low temperatures,
$T>\gamma_{e}H_{0}\hbar /k_{B}$, we obtain
$\overline{M_{z}}(\omega_{0},T)\approx1$.
From (\ref{eq.18}) we also see that
\begin{equation}\label{eq.21}
    \frac{\overline{M_{x}}(\omega_{0},T)}
    {\overline{M_{y}}(\omega_{0},T)}\approx-\frac{2a}{b}\frac{k_{B}T}{\gamma_{e}H_{0}\hbar}.
\end{equation}
The meaning of Eq. (\ref{eq.21}) is that we can rotate the magnetization's transversal
component in the plane via (cf. Fig.1).

Using (A1) we find
\begin{eqnarray}\label{eq.22}
    &&\bar{U}=-\frac{\partial}{\partial\beta}\ln\bigg(\frac{\sin(\beta H_{0})}{\beta H_{0}}\bigg) \\
    &&- \frac{\partial}{\partial \beta}\ln \bigg(1+\frac{\varepsilon^2}{2}(\delta^2+\eta^2)\frac{L(\beta H_{0})}{\beta
    H_{0}}\bigg).\nonumber
\end{eqnarray}
and for the mean energy we obtain:
\begin{eqnarray}\label{eq.23}
    &&\bar{U}=-\gamma_{e}H_{0}\hbar
     L\bigg(\frac{\gamma_{e}H_{0}\hbar}{k_{B}T}\bigg)- \\
    &&-\frac{\hbar}{2\omega_{0}}\Bigg(\frac{a^2s2}{\lambda^2}+\frac{3}{4}\frac{b^2s^2}{\lambda^2}\bigg(\frac{\gamma_{e}H_{0}\hbar}{k_{B}T}\bigg)^{2}\Bigg)L\bigg(\frac{\gamma_{e}H_{0}\hbar}{k_{B}T}\bigg)-\nonumber\\
&&-\frac{k_{B}T}{2
\omega_{0}^{2}}\Bigg(\frac{a^2s2}{\lambda^2}+\frac{1}{4}\frac{b^2s^2}{\lambda^2}\bigg(\frac{\gamma_{e}H_{0}\hbar}{k_{B}T}\bigg)^{2}\Bigg)\times
\nonumber\\&& \times
\Bigg(1-\frac{\bigg(\frac{\gamma_{e}H_{0}\hbar}{k_{B}T}\bigg)^{2}}{\sinh^{2}\bigg(\frac{\gamma_{e}H_{0}\hbar}{k_{B}T}\bigg)}\Bigg).
\nonumber
\end{eqnarray}
We can evaluate now the heat capacity
\begin{eqnarray}
%% \nonumber to remove numbering (before each equation)
   &&C_{V}=\frac{\partial\bar{U}}{\partial
   T}=k_{B}\Bigg\{\Bigg(1-\frac{\bigg(\frac{\gamma_{e}H_{0}\hbar}{k_{B}T}\bigg)^{2}}{\sinh^{2}\bigg(\frac{\gamma_{e}H_{0}\hbar}{k_{B}T}\bigg)}\Bigg)+
   \nonumber\\ &&
   +\frac{3}{4}\frac{b^2s^2}{\lambda^2}\omega_{0}\bigg(\frac{\hbar}{k_{B}T}\bigg)^{3}L\bigg(\frac{\gamma_{e}H_{0}\hbar}{k_{B}T}\bigg)-
   \\&&+
   \frac{1}{2\omega_{0}^{2}}\frac{b^2s^2}{\lambda^2}\bigg(\frac{\gamma_{e}H_{0}\hbar}{k_{B}T}\bigg)^{2}\Bigg(1-\frac{\bigg(\frac{\gamma_{e}H_{0}\hbar}{k_{B}T}\bigg)^{2}}{\sinh^{2}\bigg(\frac{\gamma_{e}H_{0}\hbar}{k_{B}T}\bigg)}\Bigg)+
   \nonumber
   \\&&+\frac{1}{\omega_{0}^{2}}\Bigg(\frac{a^2s^2}{\lambda^2}+\frac{b^2s^2}{\lambda^2}\bigg(\frac{\gamma_{e}H_{0}\hbar}{k_{B}T}\bigg)^{2}\Bigg)\times
   \nonumber \\&& \times
   \frac{\bigg(\frac{\gamma_{e}H_{0}\hbar}{k_{B}T}\bigg)^{2}}{\sinh^{2}\bigg(\frac{\gamma_{e}H_{0}\hbar}{k_{B}T}\bigg)}L\bigg(\frac{\gamma_{e}H_{0}\hbar}{k_{B}T}\bigg)\Bigg\}.\nonumber
   \end{eqnarray}
and the change of the heat capacity due to the spin torque
  \begin{eqnarray}
%% \nonumber to remove numbering (before each equation)
   &&\delta C_{V}=k_{B}\Bigg\{
   \frac{3}{4}\frac{b^2s^2}{\lambda^2}\omega_{0}\bigg(\frac{\hbar}{k_{B}T}\bigg)^{3}L\bigg(\frac{\gamma_{e}H_{0}\hbar}{k_{B}T}\bigg)-
   \\&&+
   \frac{1}{2\omega_{0}^{2}}\frac{b^2s^2}{\lambda^2}\bigg(\frac{\gamma_{e}H_{0}\hbar}{k_{B}T}\bigg)^{2}\Bigg(1-\frac{\bigg(\frac{\gamma_{e}H_{0}\hbar}{k_{B}T}\bigg)^{2}}{\sinh^{2}\bigg(\frac{\gamma_{e}H_{0}\hbar}{k_{B}T}\bigg)}\Bigg)+
   \nonumber
   \\&&+\frac{1}{\omega_{0}^{2}}\Bigg(\frac{a^2s^2}{\lambda^2}+\frac{b^2s^2}{\lambda^2}\bigg(\frac{\gamma_{e}H_{0}\hbar}{k_{B}T}\bigg)^{2}\Bigg)\times
   \nonumber \\&& \times
   \frac{\bigg(\frac{\gamma_{e}H_{0}\hbar}{k_{B}T}\bigg)^{2}}{\sinh^{2}\bigg(\frac{\gamma_{e}H_{0}\hbar}{k_{B}T}\bigg)}L\bigg(\frac{\gamma_{e}H_{0}\hbar}{k_{B}T}\bigg)\Bigg\}.\nonumber
   \end{eqnarray}

\section{Numerical results}

Let us inspect the temperature dependence of the mean values of the magnetization components using the
analytical results derived in the previous section. We note again that the analytical solutions Eq. (18), Eq. (19) contain
first and second order terms. First order terms correspond to the solution in the absence of spin torque and are valid for arbitrary
values of temperature while second order terms are defined for temperatures $T>T_{cr}$ see Eq. (20). Since we measure temperature
in units of $T_{cr}$, the solution obtained using  perturbation theory is not well-defined in the vicinity of  $T\approx 1$. Therefore,
 we expect to see a slight loss of smoothness of the magnetization curves in the vicinity around this area. However,
  our main finding of thermally activated
in-plane magnetization rotation (see Eq. (20)) is well defined for arbitrary values of the temperature.
Fig.1 shows the rotation of the in-plane component of the magnetization induced
by the change of the temperature and is plotted using Eq.(18).

\begin{figure}\label{Fig.1}
\includegraphics*[width=0.5\textwidth]{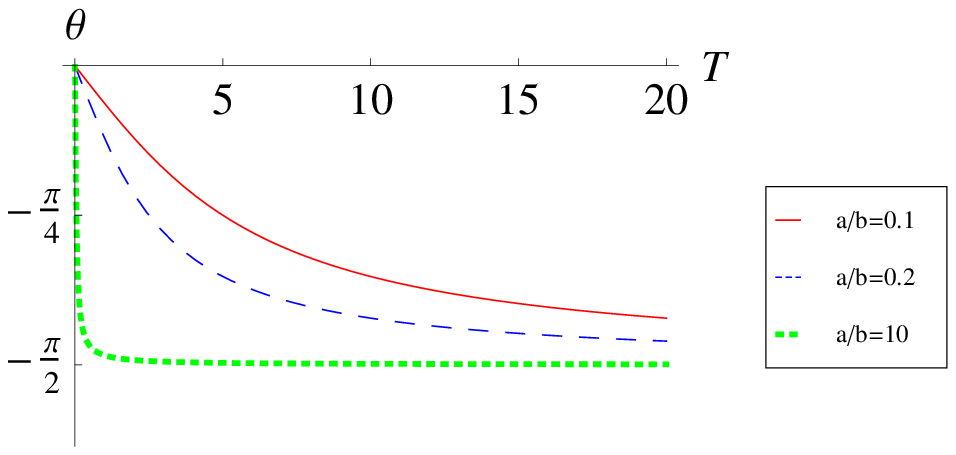}
\caption{Rotation of the magnetization in the  $xy$ plane with a
varying  ratios between the spin torque parameters $a,b$. The angle of
the rotation is defined via
$\theta(T)=\tan^{-1}\bigg(\frac{M_{x}(T)}{M_{y}(T)}\bigg)$.
Plotted using Eq.(\ref{eq.18}). The temperature unit is defined via
$\frac{\omega_{0}\hbar}{k_{B}},$ here $\omega_{0}=\gamma_{e}H_{0}$
is the Larmor precession frequency  $\omega_{0}=1$, $\lambda=1$.
The maximal  rotation angle of the magnetization
$\Delta\theta\approx\pi/2$ is reached for  the
temperature $\Delta T\approx\frac{10\omega_{0}\hbar}{k_{B}}$.}
\end{figure}

\begin{figure}\label{Fig.2}
\includegraphics*[width=0.5\textwidth]{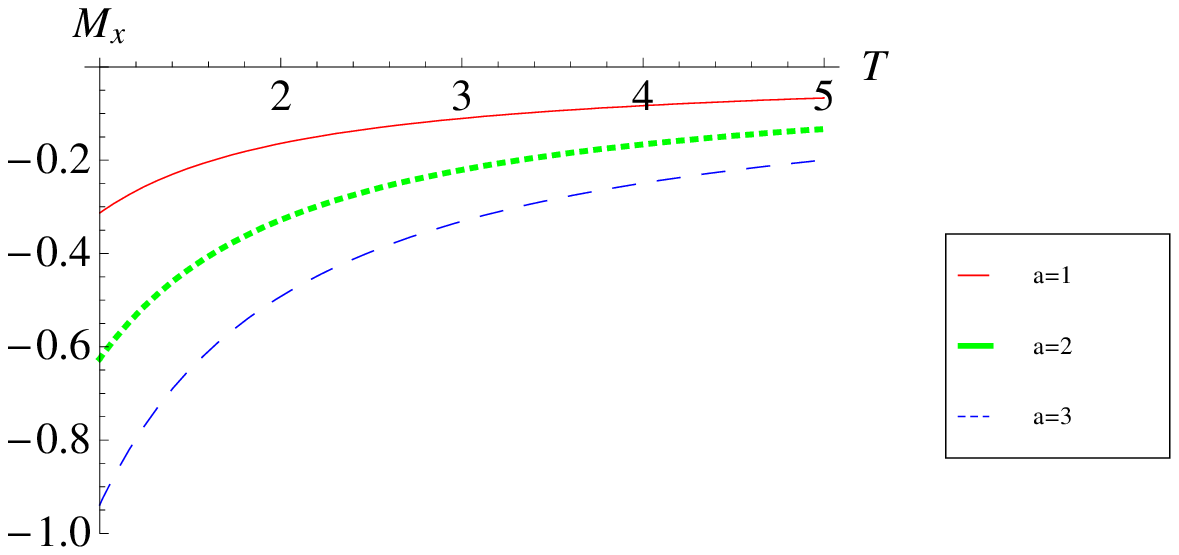}
\caption{Dependence of the magnetization component $M_{x}(T)$ on
the temperature for different values of the spin torque constant
$a$, plotted using Eq.(\ref{eq.18}). The temperature unit is defined
via $\frac{\omega_{0}\hbar}{k_{B}},$ here
$\omega_{0}=\gamma_{e}H_{0}$ is the Larmor precession frequency
$\omega_{0}=1$, $\lambda=1$. }
\end{figure}

\begin{figure}\label{Fig.3}
\includegraphics*[width=0.5\textwidth]{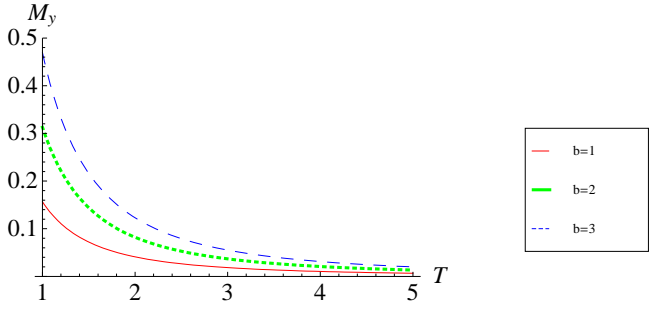}
\caption{Dependence of the magnetization component $M_{y}(T)$ on
the temperature for different values of the spin torque constant
$b$. Temperature unit is defined via
$\frac{\omega_{0}\hbar}{k_{B}},$ here $\omega_{0}=\gamma_{e}H_{0}$
is the Larmor precession frequency $\omega_{0}=1$, $\lambda=1$. }
\end{figure}

\begin{figure}\label{Fig.4}
\includegraphics*[width=0.5\textwidth]{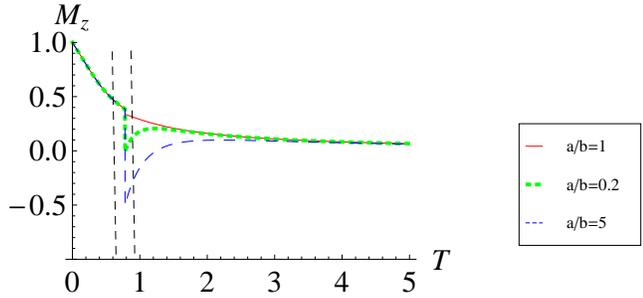}
\caption{Same as Fig.3 for  the magnetization component $M_{z}(T)$. Slight
loss of the smoothness of the magnetization curve in the vicinity $T\sim 1$ is connected to the fact
that perturbation solution is not well-defined in the area which is marked out by dashed lines .
%the temperature for different values of the spin torque constant
%$a,~~b$.Plotted using Eq.(\ref{eq.18}). Temperature unit is
%defined via $\frac{\omega_{0}\hbar}{k_{B}},$ here
%$\omega_{0}=\gamma_{e}H_{0}$ is the Larmor precession frequency
%$\omega_{0}=1$, $\lambda=1$.
 }
\end{figure}

%\newpage

Note, that the expression for the $\overline{M_{z}}(H_{0},T)$ in Eq.(\ref{eq.18})
contains two terms. The first term recovers the result
obtained without the spin torque Eq.(\ref{eq.11}) and is defined for
arbitrary temperatures. While the second term in Eq. (\ref{eq.18}) is the
contribution of the perturbation theory and therefore according to
 Eq. (\ref{eq.20}) is defined only for temperatures above
$T_{cr}$. We should take this into account when plotting
$\overline{M_{z}}(H_{0},T)$ using Eq.(\ref{eq.20}).
We see that the rotation amplitude depends on the ratio between the spin torque constants $a/b$ and for $a/b>1$
has a maximum. The temperature dependence of the mean values of the in-plane magnetization components $M_{x},~M_{y}$
is shown in Fig.2, Fig.3, and Fig.4. We see that the maximal values of  $M_{z}$ decreases with the increase of
the spin torque component $a$ (see Fig.4).

%\newpage

Now we present square components of the magnetization plotted using Eq. (19). See Fig.5, Fig.6 and Fig.7.

\begin{figure}\label{Fig.5}
\includegraphics*[width=0.5\textwidth]{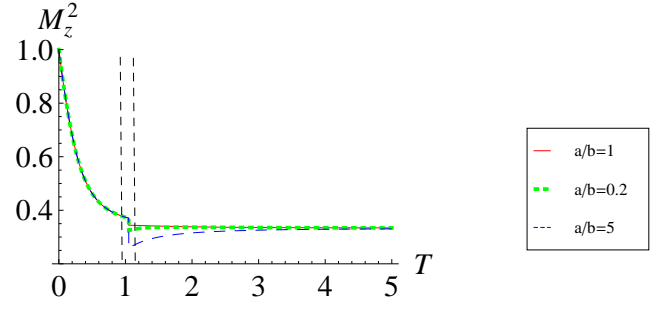}
\caption{Dependence of the magnetization component $M_{z}^{2}(T)$
on the temperature for different values of the spin torque
constant $a,~~b$, plotted using Eq.(\ref{eq.18}). Area in which perturbation theory is not well defined is marked out by two
dashed lines. Temperature unit
is given by  $\frac{\omega_{0}\hbar}{k_{B}},$  with
$\omega_{0}=\gamma_{e}H_{0}$, $\omega_{0}=1$,  and $\lambda=1$. }
%The reason for the sudden change  in $M_{z}^{2}(T)$ in the
% vicinity of $T~1$ could be troubles of the perturbation theory. }
\end{figure}

\begin{figure}\label{Fig.6}
\includegraphics*[width=0.5\textwidth]{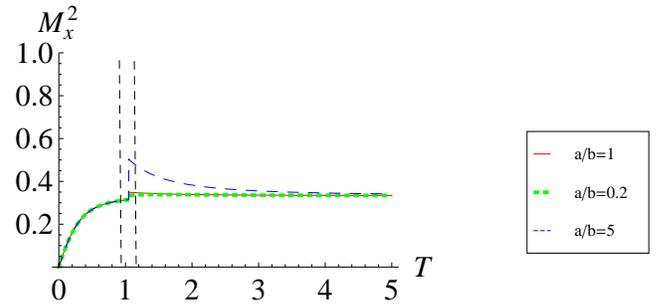}
\caption{The same as Fig.5 but for $M_{x}^{2}(T)$.
Area in which perturbation theory is not well defined is marked out by two
dashed lines.}
%
%Dependence of the magnetization component
%on the temperature for different values of the spin torque
%constant $a,~~b$.Plotted using Eq.(\ref{eq.18}). Temperature unit
%is defined via $\frac{\omega_{0}\hbar}{k_{B}},$ here
%$\omega_{0}=\gamma_{e}H_{0}$ is the Larmor precession frequency
%$\omega_{0}=1$, $\lambda=1$. The Reason for the small brake in the
%vicinity of $T\approx 1$ could be troubles of the perturbation theory. }
\end{figure}

\begin{figure}\label{Fig.7}
\includegraphics*[width=0.5\textwidth]{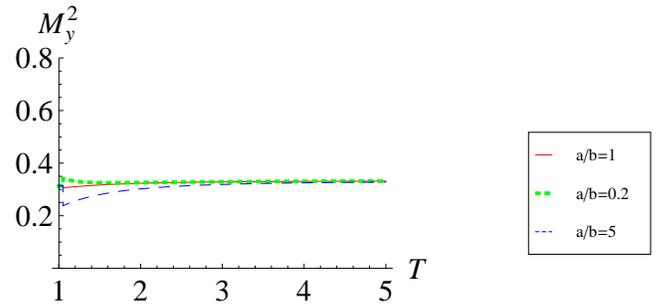}
\caption{
The same as Fig.5 but for $M_{y}^{2}(T)$.}
%Dependence of the magnetization component $M_{y}^{2}(T)$
%on the temperature for different values of the spin torque
%constant $a,~~b$.Plotted using Eq.(\ref{eq.18}). Temperature unit
%is defined via $\frac{\omega_{0}\hbar}{k_{B}},$ here
%$\omega_{0}=\gamma_{e}H_{0}$ is the Larmor precession frequency
%$\omega_{0}=1$, $\lambda=1$.}
\end{figure}

%\newpage

%\newpage

The general conclusion is that the asymmetry between the spin torque
coefficients $a,b$ has more important consequences for the thermal
rotation of the magnetization in the $xy$  plane and the mean values
of the magnetization components $\bar{M}_{x,y,z}$ however, it is less
evident for the mean values of the square of the components
$\bar{M}_{x,y,z}^{2}$. Finally, we show the temperature dependence of the dispersion for the $M_{z}$ component of the
magnetization. We see that for different ratios between the spin torque constants $a/b$, the values of the dispersion are different.
At higher temperatures all these different values merge together.

Additionally, we perform full numerical finite-temperature calculations based on the solution of the stochastic LLG equation by means of the Heun method which converges in quadratic mean to the solution interpreted in the sense of Stratonovich \cite{Kamp07}. Exact numerical
solution of the stochastic LLG equation is important since analytical results are obtained in the framework of perturbation theory
and therefore are valid for the temperatures above critical temperature $T>T_{cr}$ only. In order to observe dependence of the magnetization components on the temperature and spin torque constants, we numerically solve stochastic LLG equation Eq. (1) and generate random trajectories on the sufficiently large time interval until magnetization components after relaxation process reaches stationary regime. In the stationary regime, values of the magnetization components are time independent and depend on the temperature and spin torque parameters only. Therefore after averaging results over the ensemble of random trajectories for the magnetization components we obtain mean values which we can compare to the mean values of the magnetization
components obtained via the solution of stationary Fokker-Plank equation Eq. (8). In Fig. 9 the rotation of the magnetization denoted by the angle $\theta$ reproduces the analytical results of Eq. (21). As we see, depending on the ratio between the spin torque constants $a/b$ maximal values of the observed rotational angle $\big(\Delta\theta\big)_{max}\approx \pi /2$ is in a good quantitative and qualitative agreement with the analytical results presented in the Fig. 1.  Fig. 10 shows all three magnetization components for the chosen values of the spin current. These results are in full agreement with the analytical results depicted in Figs. 2-4, which predict a decay of the magnetization with increasing  temperatures. We have deviation between analytical and numerical results only in the area below critical temperature $T<T_{cr}\approx1$ where perturbation theory used in analytical calculations is not defined. Our full numerical results supplement for the low temperature case, i.e. for $T<1$ in dimensionless units, or in real (non-scaled) units $T<70$~[K]. The numerics can go  beyond the range of validity of the perturbation theory.
The  numerically accurate  results for the magnetization are smooth. Fig. 11 additionally presents the zero temperature equilibrium from which for certain value of the ratio $a/b$ the non-zero temperature calculations start. In particular Fig. 11 defines equilibrium ground state of the system for the zero temperature. This zero temperature ground state depends on the torque parameters. Finally, in Fig. 12 we show the effect of the magnetization rotation calculated for each time step for the averaged values of the squared projections of the magnetization, i.e. $\overline{M^2_{\mathrm{x,y,z}}}$.

\begin{figure}[h]\label{Fig.8}
\includegraphics*[width=0.5\textwidth]{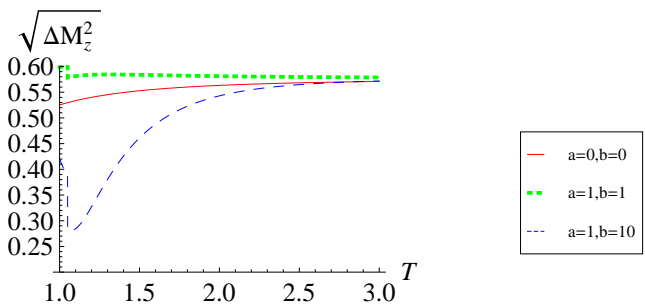}
\caption{The dispersion for the $z$ component of the
magnetization, without the spin torque $\Delta
M_{z}(T)=\sqrt{\overline{(\Delta M_{z})^2}}=\left[
1-2\frac{T}{H_{0}}L(H_{0}/T)-L^2(H_{0}/T)\right] ^{1/2}$ and in
the case of a spin torque  $\Delta
M_{z}^{ST}(T)=\sqrt{\overline{(\Delta
M_{z})^2}}=\sqrt{\overline{M^2_{z}}-\big(\overline{M_{z}}\big)^2}$.
The mean values for the  case in the presence of the torque  are
defined via Eq.(\ref{eq.18}), Eq.(\ref{eq.19}).  The temperature
unit is the same as Fig.1.}
\end{figure}

\begin{center}
   \begin{figure}[htb]
    \includegraphics[width=.32\textwidth,angle=-90]{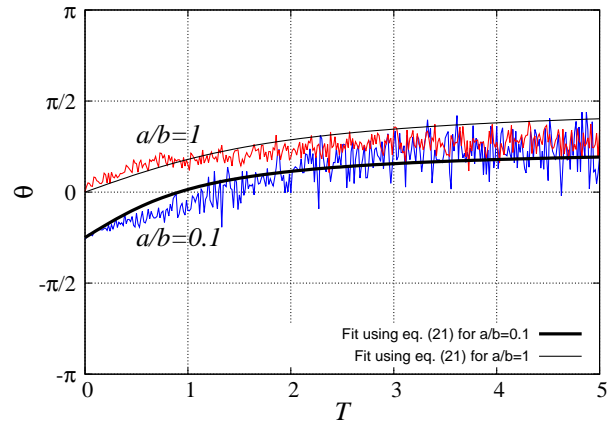}
        \caption{\label{Fig.9} (Color online) Demonstration of the magnetization rotation based on the numerical solution of the stochastic LLG. The as-obtained trajectories of the magnetization for the parameters related to Fe-Ni or Fe-Co-Ni (saturation magnetization $M_{\mathrm{S}}=1025$~[G], spin-torque $s=(s,0,0)$, external magnetic field $H_{0 \mathrm{z}}=10^4$~[Oe] and rescaled damping constant $\lambda=1$). Ensemble-averaged over 100 realizations for each time step after assuring that the magnetization reached the quasi-equilibrium scaling is done. The definition of angle $\theta(T)$ is given in the caption of Fig. 1.
        We see that numerical result fits qualitatively and quantitatively with the analytical results obtained in the framework of the perturbation theory.}
\end{figure}
\end{center}

\begin{center}
   \begin{figure*}[htb]
    \includegraphics[width=.33\textwidth,angle=-90]{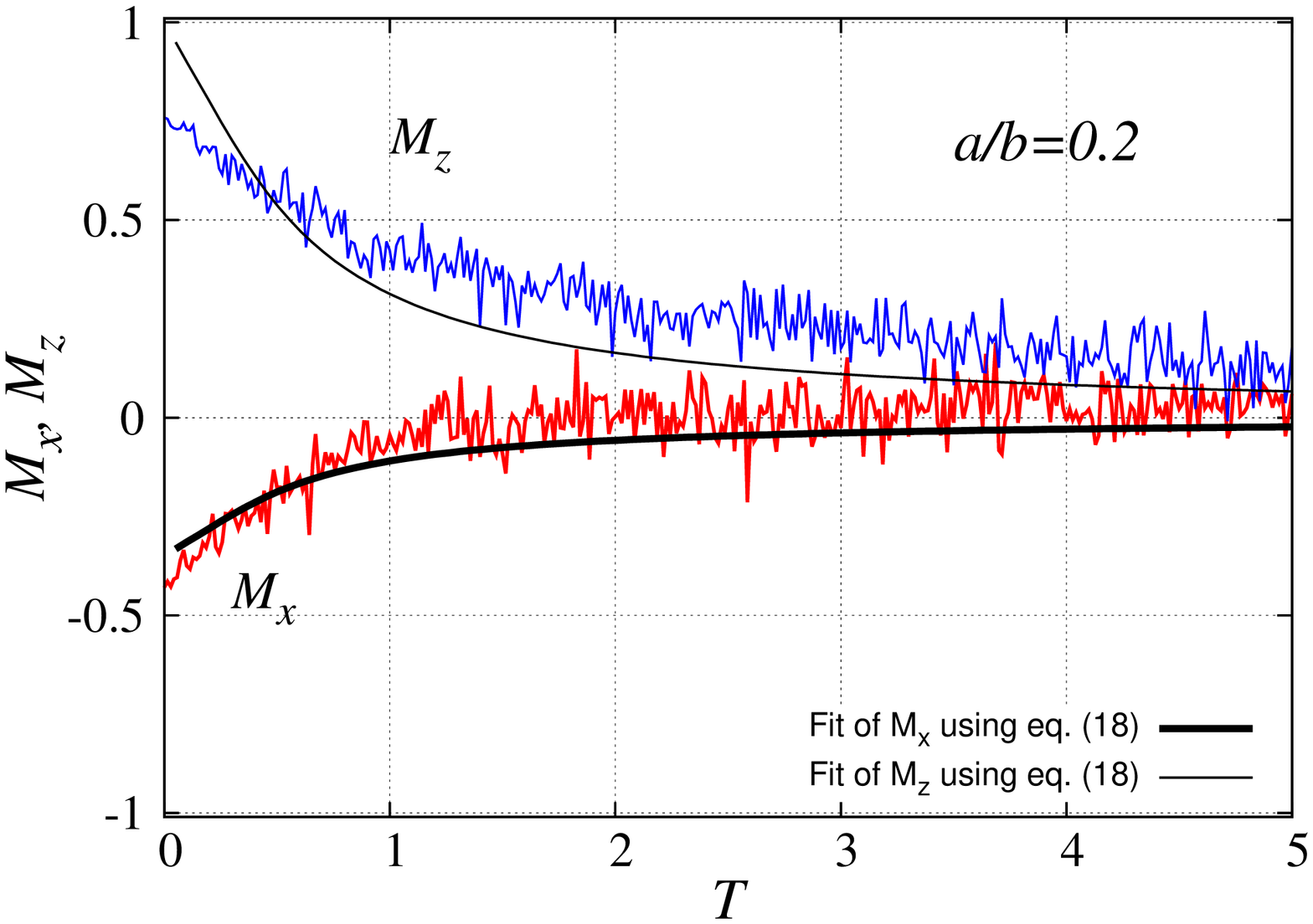}
    \includegraphics[width=.33\textwidth,angle=-90]{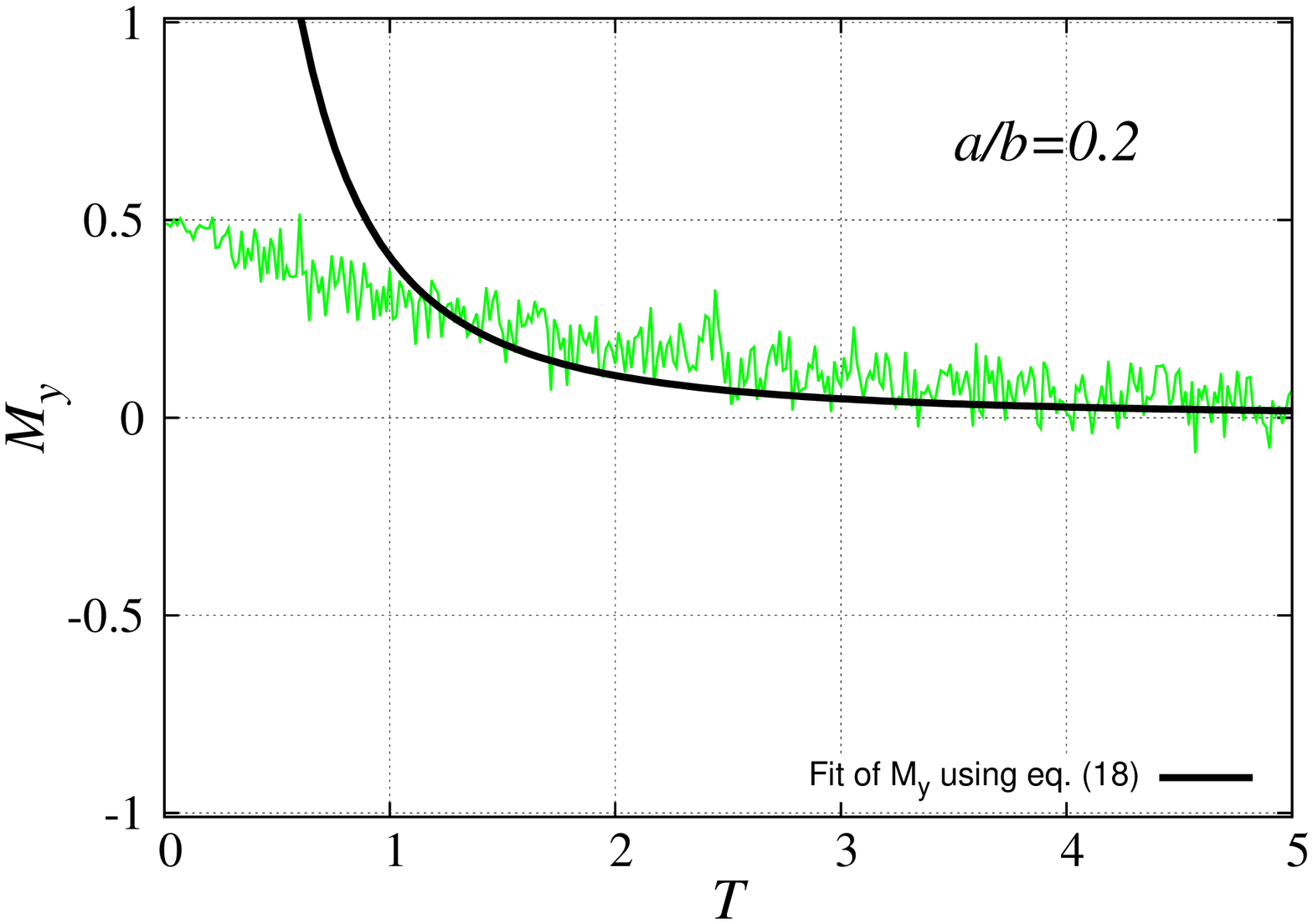}
        \caption{\label{Fig.10} (Color online) Ensemble-averaged (over 100 realizations) magnetization components calculated for the parameters listed in the caption of Fig. 9. Good agreement between numerical and analytical results is evident. We have deviation between analytical and numerical results for the component $M_{y}$ only in the area below critical temperature $T<T_{cr}\approx1$ where perturbation theory used in analytical calculations is not defined. Our full numerical results supplement for the low temperature case, i.e. for $T<1$ in dimensionless units, or in real (non-scaled) units $T<70$~[K].}
\end{figure*}
\end{center}

\begin{center}
   \begin{figure*}[htb]
    \includegraphics[width=.92\textwidth]{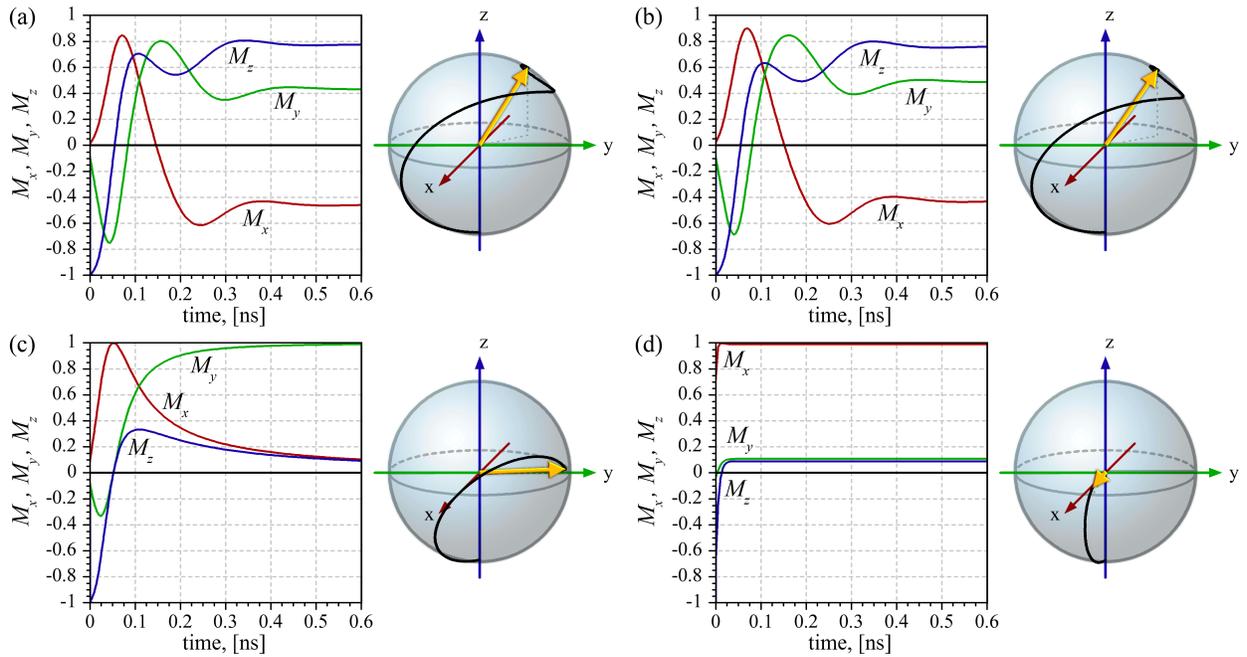}
        \caption{\label{Fig.11} (Color online) Illustration of the relaxation of the magnetization from the initially chosen arbitrary states to the zero temperature equilibrium state. In case of zero temperature relaxation of the magnetization vector is connected to the phenomenological damping constant $\lambda$. As we see due to the spin torque terms transversal components of the magnetization vector are different from the zero in the equilibrium.  Initial state is chosen as $M_{\mathrm{z}}(t=0)=-1$. a) $a/b=0.1$, b) $a/b=0.2$, c) $a/b=1$, d) $a/b=10$; other parameters are as those listed in the caption of Fig. 9.}
\end{figure*}
\end{center}

\begin{center}
   \begin{figure}[htb]
    \includegraphics[width=.29\textwidth]{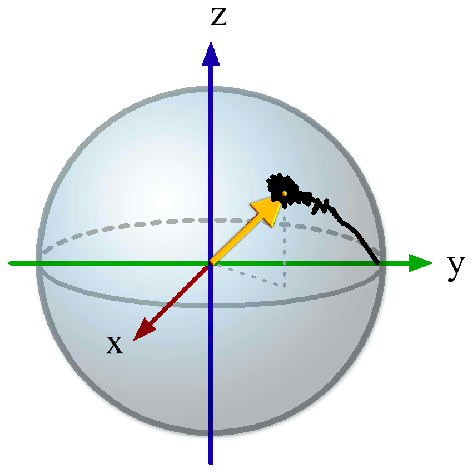}
        \caption{\label{Fig.12} (Color online) Rotation of the magnetization as a function of temperature ($T\in [0:7]$) calculated from trajectories averaged for each temperature at quasi-equilibrium after relaxation time $\tau_{\mathrm{rel}}\approx \tau_{\mathrm{precess}}/\lambda$. Number of averaging is $100$, $a/b=1$. Initial magnetization state is $\{M_{\mathrm{x}},M_{\mathrm{y}},M_{\mathrm{z}}\}(t=0)=\{0,1,0\}$.}
\end{figure}
\end{center}

\section{Conclusion}

In this work we studied the  thermally assisted  spin-torque  and its influence on the magnetization
dynamics in a thin permalloy film and presented results for
Fe$_{25}$Co$_{25}$Ni$_{50}$. We found that the spin torque term
leads to  nontrivial dynamical effects in the finite temperature  magnetization dynamics.
Assuming the spin torque terms to be   small compared to the
Larmor precessional   term  we developed a  perturbational approach
to the Fokker-Planck equation and obtained
analytical expression for the distribution function including the spin torque terms.
In particular, we proved that the spin torque term leads to the formation of a non-vanishing
in-plane magnetization component. The ratio between
the mean values of the components $\overline{M_{x}}$ and $\overline{M_{y}}$ defines
the orientation of the in-plane magnetization vector
$\frac{\overline{M_{x}}(\omega_{0},T)}
{\overline{M_{y}}(\omega_{0},T)}\approx-\frac{2a}{b}\frac{k_{B}T}{\gamma_{e}H_{0}\hbar}$.
We find  that the
orientation of the in-plane magnetization depends on the ratio between the spin torque
constants $a/b$ and between the temperature and the amplitude of the external magnetic field
$T/H_{0}$. Therefore,  changing  the temperature leads to a  thermally induced
rotation of the in-plane magnetization vector. We name this as
"thermally activated in-plane magnetization rotation". We found that if from the two
spin torque terms
$\varepsilon b\big[\vec{M},\vec{s}\big],
~~~ \varepsilon a\big[\vec{M}\big[\vec{M},\vec{s}\big]\big]$
the last term is the dominant one $a>b$ the effect of the
thermally activated in-plane magnetization rotation is enhanced (cf. Fig.1) .

\section*{Acknowledgements}
Financial support by the Deutsche Forschungsgemeinschaft (DFG) through SFB 762 and SU 690/1-1 and contract BE 2161/5-1 is gratefully acknowledged. This work is supported by the National Science Center in Poland as a research project in years 2011 - 2014, and  CONACYT of Mexico (Basic Science Project No. 129269).

\section{Appendix}

We use the partition function
\begin{eqnarray}\label{eq.A1}
    Z(\alpha)=4\pi\frac{\sinh
    \alpha}{\alpha}\bigg(1+\frac{\varepsilon^2(\delta^2+\eta^2)}{2}\frac{L(\alpha)}{\alpha}\bigg).
    \nonumber ~~~~(A1)
\end{eqnarray}
Here we give some expressions for  the components of the magnetization
\begin{eqnarray}
    \overline{M_{x}}=\varepsilon\delta \frac{L(\beta H_{0})}{\beta H_{0}}\frac{1}{\bigg(1+\frac{\varepsilon^2(\delta^2+\eta^2)}{2}\frac{L(\beta H_{0})}{\beta H_{0}}\bigg)}\nonumber\\
   \approx \varepsilon\delta \frac{L(\beta H_{0})}{\beta H_{0}}\bigg(1-\frac{\varepsilon^2(\delta^2+\eta^2)}{2}\frac{L(\beta H_{0})}{\beta H_{0}}\bigg),\nonumber
\end{eqnarray}
\begin{eqnarray}\label{eq.A2}
    \overline{M_{y}}=-\varepsilon\eta \frac{L(\beta H_{0})}{\beta H_{0}}\frac{1}{\bigg(1+\frac{\varepsilon^2(\delta^2+\eta^2)}{2}\frac{L(\beta H_{0})}{\beta H_{0}}\bigg)}\nonumber\\
   \approx -\varepsilon\eta \frac{L(\beta H_{0})}{\beta H_{0}}
   \bigg(1-\frac{\varepsilon^2(\delta^2+\eta^2)}{2}\frac{L(\beta H_{0})}{\beta H_{0}}\bigg),\nonumber~~(A2)
\end{eqnarray}
\begin{eqnarray}
   \overline{M_{z}}=\frac{L(\beta H_{0})}
   {\bigg(1+\frac{\varepsilon^2(\delta^2+\eta^2)}{2}\frac{L(\beta H_{0})}{\beta H_{0}}\bigg)}\nonumber\\
   +\frac{\frac{\varepsilon^2(\delta^2+\eta^2)}{2}\frac{1}{\beta H_{0}}\bigg(1-3\frac{L(\beta H_{0})}{\beta H_{0}}\bigg)}
   {\bigg(1+\frac{\varepsilon^2(\delta^2+\eta^2)}{2}\frac{L(\beta H_{0})}{\beta H_{0}}\bigg)}\nonumber \\
   \approx L(\beta H_{0})\bigg(1-\frac{\varepsilon^2(\delta^2+\eta^2)}{2}\frac{L(\beta H_{0})}{\beta H_{0}}\bigg)\nonumber\\
   +\frac{\varepsilon^2(\delta^2+\eta^2)}{2\beta H_{0}}\bigg(1-\frac{3L(\beta H_{0})}
   {\beta H_{0}}\bigg)\nonumber\\
   \times \bigg(1-\frac{\varepsilon^2(\delta^2+\eta^2)}{2}\frac{L(\beta H_{0})}{\beta H_{0}}\bigg)\nonumber.
\end{eqnarray}
For the modulus squares of the magnetization components we find
\begin{eqnarray}
    \overline{M_{x}^2}=\frac{\frac{L(\beta H_{0})}{\beta H_{0}}}
    {\bigg(1+\frac{\varepsilon^2(\delta^2+\eta^2)}{2}\frac{L(\beta H)}{\beta H_{0}}\bigg)}\nonumber\\
    +\frac{\frac{\varepsilon^2(3\delta^2+\eta^2)}{2(\beta H)^2}\bigg(1-\frac{3L(\beta H_{0})}{\beta H_{0}}\bigg)}{\bigg(1+\frac{\varepsilon^2(\delta^2+\eta^2)}{2}\frac{L(\beta H_{0})}{\beta H_{0}}\bigg)}\nonumber
\end{eqnarray}
\begin{eqnarray}
    \overline{M_{y}^2}= \frac{\frac{L(\beta H_{0})}{\beta H_{0}}}
    {\bigg(1+\frac{\varepsilon^2(\delta^2+\eta^2)}{2}\frac{L(\beta H_{0})}{\beta H_{0}}\bigg)}+
    \frac{\frac{\varepsilon^2(\delta^2+3\eta^2)}{2(\beta H_{0})^2}\bigg(1-\frac{3L(\beta H_{0})}
    {\beta H_{0}}\bigg)}{\bigg(1+\frac{\eta^2(\delta^2+\eta^2)}{2}\frac{L(\beta H_{0})}{\beta H_{0}}\bigg)},
    \nonumber
\end{eqnarray}
\begin{eqnarray}
    \overline{M_{z}^2}= \frac{1-2\frac{L(\beta H)}{\beta H_{0}}}
    {\bigg(1+\frac{\varepsilon^2(\delta^2+\eta^2)}{2}\frac{L(\beta H_{0})}{\beta H_{0}}\bigg)}\nonumber\\
    +\frac{\frac{\varepsilon^2(\delta^2+\eta^2)}{2}\Bigg(\frac{4}
    {(\beta H_{0})^2}\bigg(\frac{3L(\beta H_{0})}{\beta H_{0}}-1\bigg)+\frac{L(\beta H_{0})}
    {\beta H_{0}}\Bigg)}{\bigg(1+\frac{\varepsilon^2(\delta^2+\eta^2)}{2}\frac{L(\beta H_{0})}
    {\beta H_{0}}\bigg)},~~~~~(A3)\nonumber
\end{eqnarray}
where $\alpha=\beta H_{0}=\gamma_{e}H_{0}\hbar /k_{B}T$.

\end{document}